\documentclass[aps,nofootinbib]{revtex4}
\usepackage{amstext,amsmath,amssymb,amsfonts,bbm}
\usepackage[latin1]{inputenc}
\usepackage{fancyhdr}
\usepackage[dvips]{graphicx}
\usepackage{epsfig}
\usepackage{hyperref}

\def\bra{\langle} \def\ket{\rangle}
\def\f{\frac}
\newcommand{\SU}{\mathrm{SU}}

\newcommand{\SO}{\mathrm{SO}}

\newcommand{\U}{\mathrm{U}}
\newcommand{\lalg}[1]{\mathfrak{#1}}
\newcommand{\su}{\lalg{su}} 
\renewcommand{\u}{\lalg{u}} 
 
 \newcommand{\spin}{\lalg{spin}}

\def\mone{^{-1}}
\def\tl{\widetilde}
\def\pp{\partial}
 \def\arr{\rightarrow}
\def\eps{\epsilon}
\def\vareps{\varepsilon}

\newcommand{\tr}{\mathrm{tr}}
\newcommand{\Tr}{\mathrm{Tr}}

\def\N{{\mathbbm N}}
\def\Z{{\mathbbm Z}}
\def\R{{\mathbbm R}}
\def\id{\mathrm{id}}

\def\beq{\begin{equation}}
\def\ee{\end{equation}}
\def\bes{\begin{eqnarray}}
\def\ees{\end{eqnarray}}

\def\l({\left(}
\def\r){\right)}
\def\lv{\lvert}
\def\rv{\rvert}

\begin{document}
\title{{\large From lattice BF gauge theory to area-angle Regge calculus}}
\author{{\bf Valentin Bonzom}\email{valentin.bonzom@ens-lyon.fr}}
\affiliation{Centre de Physique Th\'eorique, CNRS-UMR 6207, Luminy Case 907, 13288 Marseille, France EU,}
\affiliation{Laboratoire de Physique, ENS Lyon, CNRS-UMR 5672, 46 All\'ee d'Italie, 69007 Lyon, France EU}

\begin{abstract}

We consider Riemannian 4d BF lattice gauge theory, on a triangulation of spacetime. Introducing the simplicity constraints which turn BF theory into simplicial gravity, some geometric quantities of Regge calculus, areas, and 3d and 4d dihedral angles, are identified. The parallel transport conditions are taken care of to ensure a consistent gluing of simplices. We show that these gluing relations, together with the simplicity constraints, contain the constraints of area-angle Regge calculus in a simple way, via the group structure of the underlying BF gauge theory. This provides a precise road from constrained BF theory to area-angle Regge calculus. Doing so, a framework combining variables of lattice BF theory and Regge calculus is built. The action takes a form {\it \`a la Regge} and includes the contribution of the Immirzi parameter. In the absence of simplicity constraints, the standard spin foam model for BF theory is recovered. Insertions of local observables are investigated, leading to Casimir insertions for areas and 6j-symbols for 3d angles. The present formulation is argued to be suitable for deriving spin foam models from discrete path integrals.

\end{abstract}
\maketitle

\section*{Introduction}

Lattice approaches to describe gravity take several forms. Among them, Regge calculus and spin foams are expected to be intimately related. Both are standardly built with a triangulation of spacetime. The idea of Regge calculus \cite{regge} is to concentrate curvature around $(d-2)$-simplices (triangles in four dimensions). In its original from, the geometric variables are the squared lengths of edges. The dihedral angles (the angles between adjacent tetrahedra) are functions of the edge lengths, and curvature is measured by the deficit angle around each triangle, which is basically the sum of the dihedral angles at simplices sharing that triangle. Regge calculus has been revived in the eighties (see \cite{reviewRC} for reviews and references therein) and several alternative formulations have been discussed, in connection with the developments of loop quantum gravity and spin foams: first order Regge calculus \cite{barrett first order}, using triangle areas instead of edge lengths (\cite{makela} for instance), and recently using areas and 3d dihedral angles \cite{area-angleRC}.

As for spin foams, or rather sums over spin foams, they correspond to a way of writing transition amplitudes between spin network states in gauge theories \cite{baez def SF}, \cite{reisenberger worldsheet}. A spin foam is a colored two-complex joining two spin networks. For a single two-complex which may be seen as dual to a triangulation, the sum over spin foams is a state-sum model whose data come from the representation theory of some algebraic structure (typically a Lie group). We thus expect to derive spin foam models from lattice gauge theories \cite{conrady}, exactly as in 3d quantum gravity \cite{PR1} and lattice Yang-Mills theory \cite{drouffe zuber}. This background independent framework is particularly well designed for topological BF theories \cite{baez BF}. Spin foam models for quantum gravity take advantage of a reformulation of general relativity as a constrained BF theory, Plebanski theory \cite{plebanski}. The so-called simplicity constraints ensure that the field $B$ comes from a frame field \cite{simplicity}.

On the lattice, the bivectors, discretizing the field $B$ on triangles, thus contain the geometric quantities, such as tetrahedron and 4-simplex volumes and in particular areas and 3d angles. Dittrich and Speziale \cite{area-angleRC} have argued that area-angle Regge calculus is the natural Regge discretization of Plebanski theory. Although appealing, their porposal has some practical drawbacks: the expression of the (4d) dihedral angles in terms of the 3d angles takes a complicated form, and so do the simplicity constraints.

We show in this paper that all these expressions are naturally encoded, in a simple way, through the group structure of lattice Plebanski theory. This is done by formulating cross-simplicity and the rules for parallelly transporting bivectors using only group variables. We then solve the parallel transport conditions for the holonomies in terms of the bivectors and additional angles. The latter are recognized as being the dihedral angles and can be seen as a discrete $\u(1)\oplus\u(1)$ connection for an additional gauge symmetry.

Another interrogation in spin foams and Regge calculus is the role of the Immirzi parameter. More generally, in quantum gravity literature, the reader often finds himself in view of different interpretations. While some authors bring forward a topological interpretation of the Immirzi parameter (see \cite{mercuri}, and references therein), its presence turns out to be crucial in loop quantum gravity to define the quantum geometry. It also plays an important role in the EPRL \cite{epr} and FK$\gamma$ \cite{fk}, \cite{conrady1} spin foam models. The recent semi-classical analyses \cite{conrady2}, \cite{barrett fairbairn} show that it disappears from the action at the critical points, in agreement with the continuum. However, as far as we know, it has not been introduced in any form of the Regge action. We here fill the first stage of that gap, proposing an action (in fact, two, which are quite similar) which includes the Immirzi parameter. Due to the parallel transport relations and cross-simplicity, it is a function of only areas and dihedral angles, and corresponds to the compactified Regge action \cite{caselle}, distorted by the Immirzi parameter. Similarly to the above-mentioned saddle point analyses, the Immirzi dependence vanishes on-shell.

We thus arrive at an improved setting which mixes the advantages of both lattice BF theory and Regge calculus. The geometric quantities are clearly identified. The formulation using group elements enables to translate their insertions in the path integral as local observables into insertions in the sum over spin foams. For pure $\SU(2)$ BF theory, we show that insertions of areas and 3d angles respectively translate into insertions of Casimirs on triangles and of 6j-symbols on tetrahedra.

An underlying motivation of this work is to find a framework allowing to derive the known spin foam models from the discretized path integral of constrained BF theory. The main approach to build spin foam models for quantum gravity relies on the geometric quantization of the tetrahedron classical phase space. This important idea has been developped in \cite{quantum tet} by Barbieri, Baez and Barrett. The EPRL model then imposes the constraints at the quantum level, via equations on Casimir operators. The derivation of the FK$\gamma$ model is different in spirit, but is also based on the quantization of the tetrahedron phase space, using coherent states \cite{etera simone}. If these works have given a clear picture of cross-simplicity using the geometric quantization of a single tetrahedron, the issue of gluing adjacent tetrahedra is less transparent. It would be interesting to directly derive these models from a (constrained) Lagrangian approach. Several approaches have been recently developped in that direction. Baratin, Flori and Thiemann \cite{thiemann holst} have started from the Holst action, thus without need for simplicity, and used a different spacetime discretization. It would certainly be interesting to relate their results to the standard framework. Conrady and Freidel \cite{conrady1} have given a path integral representation of the FK$\gamma$ and EPRL models. It has nice geometric features, which have enabled the semi-classical analysis \cite{conrady2}. It is however, in our view, specifically designed for the new spin foam models, and we would like to have a better control of the imposition of the constraints and of the gluing of tetrahedra. Another programme has been initiated in \cite{lagrangian BC}, in which a Lagrangian derivation of the Barrett-Crane spin foam model \cite{bc} is given. It shows that the main problem of that model is not the implementation of cross-simplicity, but rather the gluing process between adjacent tetrahedra. Indeed, the method allowing to recover the BC model is that used to generate spin foam models for {\it unconstrained} BF-like theories \cite{class actions SF}, and its equations of motion fail to yield the expected rules for parallelly transporting bivectors in the BC context. The present framework enables to choose and precisely control the way these gluing relations are taken into account in the path integral. Since only group elements are used as configuration variables (except areas), this formulation is particularly convenient to derive spin foam models using Fourier expansions on the groups.

The paper is organized as follows. In section \ref{Lattice BF gauge theory}, we review the standard lattice approach to BF theory, and solve the parallel transport relations for the holonomies. In section \ref{area-angle RC}, we extract the geometric quantities of area-angle Regge calculus. It is shown that simplicity together with parallel transport naturally contain the expressions of the (4d) dihedral angles and of the constraints of area-angle Regge calculus. We also study the action and the role of the Immirzi parameter. It is shown in section \ref{SF BF} that the present setting reproduces the standard spin foam model for pure BF theory, and we compute the insertions of areas and 3d angles.

During the completion of this work, we have learnt that Barrett {\it et al} \cite{barrett fairbairn} have developped a similar analysis, at least to extract the deficit angles, but in a quite different context. They are indeed interested in the semi-classical behaviour of the EPRL model, which turns out to admit a Regge form. Our point of view is here that the definition of the discretized functional integral from constrained BF theory should itself have an interpretation {\it \`a la Regge}, since both are basically lattice gravity. It turns out that this idea has also been adopted by Oriti in \cite{oriti}, from the group field theory point of view, with a similar emphasis on parallel transport relations.

We are all along concerned with Riemannian gravity, so that the relevant structure group is Spin(4). The Lorentzian version of the present analysis will be studied elsewhere.

\section{Lattice BF gauge theory} \label{Lattice BF gauge theory}

\subsection{The standard discretization}

Let $P$ be a principal $G$-bundle over a 4d smooth manifold $M$, spacetime, $G=\SU(2)$ or Spin(4). Consider a connection $A$ over $P$, which will be locally seen, as usual, as a 1-form taking values in the Lie algebra $\lalg{g}$ of $G$ ($A^{IJ}$ where $I,J=0,1,2,3$ are Euclidean 4d indices, and $A^i$, $i=1,2,3$ in the case of $\SU(2)$). The action of the topological quantum field theory called BF \cite{horowitz}, \cite{blau} is built with a $\lalg{g}$-valued 2-form field $B$, transforming under the adjoint representation of $G$:
\beq
S_{\mathrm{BF}} = \int_M \Tr\Bigl(B\wedge F(A)\Bigr)
\ee
where $F(A)=dA+\f{1}{2}[A,A]$ is the curvature of $A$.

The action is gauge invariant and the equations of motion are:
\begin{align}
d_A B &= 0 \label{daB}\\
F(A) &= 0 \label{flatness}
\end{align}
where $d_A = d + [A,\cdot]$ is the covariant derivative. Thus $B$ can be seen as a Lagrange multiplier imposing $A$ to be flat. Moreover, the field $B$ can be completely gauged away due to an additional symmetry: $B' = B + d_A \phi$ for any $\lalg{g}$-valued 1-form $\phi$, while $A$ is unchanged. The theory has thus no local degrees of freedom.


To quantize BF theory, we proceed in a standard way by first discretizing the variables. To fill the gap from lattice BF theory to Regge calculus via the Plebanski constraints, we choose a setting well-adapted to Regge calculus. Consider a simplicial decomposition of spacetime and ask for local frames on 3-simplices, i.e. tetrahedra, and on 4-simplices. This means that curvature is concentrated around 2-simplices, i.e. triangles. We will use the dual skeleton to the triangulation in an almost systematic way. In the dual picture, triangles are dual to faces, both denoted $f$, tetrahedra to edges, both denoted $t$ and 4-simplices to points denoted $v$. The boundary of a dual face is made of the edges and vertices respectively dual to the tetrahedra and 4-simplices sharing $f$. The orientations of tetrahedra and triangles induce orientations for dual edges and dual faces.

As far as BF theory is concerned, the case $G=$Spin(4) is just two copies of the $\SU(2)$ case, so that we focus here on $G=\SU(2)$. The connection is discretized like in usual lattice gauge theory: we consider $\SU(2)$ elements $g_{vt}$ which allow for parallel transport between local frames. In the dual skeleton picture, the two ends of a dual edge $t$ correspond to the two 4-simplices sharing the tetrahedron $t$. Thus, a dual edge is attached two group elements $g_{vt}$, one for each end. The curvature around a triangle $f$ is thus measured by the oriented product of these group elements all along the boundary of the dual face, starting at a base point (reference frame) $v^\star$, $g_f(v^\star) = 
\prod_{t\subset \pp f} g_{vt} g_{v't}\mone$ if $v$, $v'$ are source and target vertices for each dual edge $t$. The flatness imposed by the e.o.m. then reads: $g_f(v)=\id$\footnotemark for each $f$.

\footnotetext{In fact, due to the use of group elements in the action, and not Lie algebra elements, the action, given below, only catches the projection of $g_f$ onto the Pauli matrices so that only the sine of the class angle of $g_f$ is restricted to be zero. The class angle can thus be 0 or $2\pi$, i.e. $g_f = \pm\id$}

The field $B$, as a 2-form is discretized on the triangles of the triangulation, in a given frame for each of them, $b_f(v)$ or $b_f(t)\in\su(2)$. Expressions of $b_f$ in different frames are naturally related by parallel transport along the boundary of $f$. It simply results from the discretisation of the e.o.m. \eqref{daB}, integrated along dual edges between different tetrahedra or 4-simplices.

$\SU(2)$ gauge transformations change the local frames on tetrahedra and 4-simplices. A gauge transformation $k$ is a family of group elements $\{k_t,k_v\}$ acting by:
\begin{align} \label{gauge transfo}
&k\,\triangleright g_{vt} = k_v\,g_{vt}\,k_t\mone \\
&k\,\triangleright b_f(t) = k_t\,b_f(t)\,k_t\mone
\end{align}
The gauge invariant action takes the form of a sum over triangles:
\beq
S_{\mathrm{BF}}\bigl(b_f(t),g_{vt}\bigr) = \sum_f \tr\bigl(b_f(t)\,g_f(t)\bigr)
\ee
which is obviously independent of the choice of the base point for each triangle. Because of the simple rules of parallel transport for the $b_f(t)$s, one can choose as basic variables the elements $g_{vt}$ and only one $b_f(t)$ per face.

Things are not so simple when introducing the simplicity constraints, and it is then convenient, in order to solve them within each tetrahedron, to start with independent $b_{ft}$ in the frame of each tetrahedron sharing $f$. It is as if the triangulation were broken up into a disjoint union of tetrahedra. While solving simplicity within tetrahedron, some gluing relations are necessary to stick tetrahedra together and carry the information about parallelly transporting $b_f$. The idea is thus to use a measure such as:
\beq
\prod_{(t,v)}dg_{vt}\ \prod_{(f,t)} d^3b_{ft}\ \prod_{(f,v)}\delta\bigl(b_{ft}- g_{tt'}\,b_{ft'}\,g_{tt'}\mone\bigr) \label{su2 gluing}
\ee
in $\SU(2)$ matrix notation, and with: 
\beq
g_{tt'} = g_{vt}\mone\,g_{vt'}
\ee
standing for the parallel transport between two adjacent tetrahedra $t$ and $t'$ through the 4-simplex $v$, along the boundary of the dual face $f$. $dg$ is the $\SU(2)$ Haar measure and the measure $d^3b$ is to be precised in the following. The gluing relations take place at each pair $(f,v)$, often called wedge, since two tetrahedra sharing a triangle $f$ lie in a common 4-simplex $v$, $f$ and $v$ identifying them.

The structure group being $G=\SU(2)$, the norm and the direction of $b_{ft}$ have different roles. We will also see that the simplicity constraints assign them different roles in order to reconstruct metricity. First, it is clear that the norm of $b_{ft}$ and that of $b_{ft'}$ are equal, $A_{ft}=A_{ft'}\equiv A_f$: it is a $\SU(2)$ gauge invariant quantity. Let us use the following parametrisation for each $b_{ft}$ as a 2 by 2 matrix:
\beq \label{b parametrisation}
b_{ft} = \f{i}{2}A_f\,\hat{b}_{ft}\cdot\vec{\sigma} ,\qquad\qquad \hat{b}_{ft}\cdot\vec{\sigma} = n_{ft}\,\sigma_z\, n_{ft}\mone
\ee
with $n_{ft}\in\SU(2)$. $\sigma_z$ is the standard Pauli matrix diag$(1,-1)$, and $\vec{\sigma}$ is the 3-vector whose components are the Pauli matrices. $A_f$ stands for the norm of each $b_{ft}$, $A_f^2 = -2\tr(b_{ft}^2)$, and is clearly independent of any local frame. For reasons which will become obvious in the sequel, we will also call it the area of $f$. Notice that we can equivalently consider $A_f\in\R$ or $\R_+$, since a change of sign of the area can be reabsorbed into the directions $\hat{b}_{ft}$ without changing the action. The direction $\hat{b}_{ft}\in S^2$ of $b_{ft}$ is encoded in the group element $n_{ft}$. However only two parameters of $n_{ft}$ are relevant since $b_{ft}$ is invariant under the right action of the $\U(1)$ subgroup generated by $\sigma_z$, $n_{ft}\rightarrow n_{ft}\,e^{-\f{i}{2}\theta\sigma_z}$. It is clear in the Euler parametrisation of $\SU(2)$, $g=e^{-\f{i}{2}\alpha\sigma_z} e^{-\f{i}{2}\beta\sigma_y} e^{-\f{i}{2}\gamma\sigma_z}$, that the angle $\gamma$ is an ambiguity in the definition, which should be integrated out. This allows to rewrite the measure using the $\U(1)$ and $\SU(2)$ Haar measures :
\beq \label{gluing measure} \begin{split}
&\prod_f d\mu\bigl(A_f\bigr)\ \prod_{(f,t)} d^2\hat{b}_{ft}\ \prod_{(f,v)} \delta^{(2)}\bigl(\hat{b}_{ft}-R\bigl(g_{tt'}\bigr)\hat{b}_{ft'}\bigr) \\ 
&\arr \quad \prod_f d\mu\bigl(A_f\bigr)\ \prod_{(f,t)} dn_{ft}\ \prod_{(f,v)} \int_0^{4\pi}d\theta_{fv}\,\delta_{\SU(2)}\Bigl(n_{ft}\mone\ g_{tt'}\,n_{ft'}\ e^{\f{i}{2}\theta_{fv}\sigma_z}\Bigr) \end{split}
\ee
$R$ simply denotes the vector representation of $\SU(2)$. $dn_{ft}$ is the $\SU(2)$ Haar measure and on the right side of this equation, it is understood that the Euler angle $\gamma$ of each $n_{ft}$ does not play any role since it can be reabsorbed into the angles $\theta_{fv}$ which are integrated out in the measure. We are not interested in the precise correspondence, including Jacobians, between the two above expressions, since we look for the simplest and most natural measure with regard to lattice gauge theory, thus using group elements and the corresponding Haar measures as much as possible. The remaining ambiguity is the measure $d\mu(A)$, which will be fixed to be the Lebesgue measure on $\R$ by requiring to recover the standard spin foam model for $\SU(2)$ BF theory (see section \ref{SF BF}).

\subsection{BF geometry}

Up to now, the configuration variables are the group elements $g_{vt}$, the norms $A_f$ and the directions $\hat{b}_{ft}$, i.e. unit 3-vectors which can be encoded in the group elements $n_{ft}$ up to right $\U(1)$ multiplication. In particular, the action is a function of $g_{vt}$, $A_f$ and $\hat{b}_{ft}$. The key idea is now to consider the three angles of each $n_{ft}$ as classical configuration variables, together with the angles $\theta_{fv}$ which are integrated out to define the measure in \eqref{gluing measure}. In order to extract the dihedral angles, we need to solve the parallel transport relations \eqref{su2 gluing} for the holonomies $g_{tt'}$ in terms of the bivectors. This has been done implicitly in \eqref{gluing measure}. Given an assignment of elements $g_{vt}$ for parallel transport and $n_{ft}$ for each triangle in each tetrahedron, the gluing condition between adjacent tetrahedra implies that there exist angles $\theta_{fv}\in[0,4\pi)$ such that:
\beq \label{gluing n}
n_{ft} = g_{tt'}\,n_{ft'}\,e^{\f{i}{2}\theta_{fv}\sigma_z}
\ee
which is precisely the content of the delta functions in \eqref{gluing measure}. As emphasized, the Euler angles $\gamma$ of the variables $n_{ft}$ are unphysical since they do not apear in the expression of $\hat{b}_{ft}$. However, it seems that they do play a role in \eqref{gluing n}. The important point is in fact that $g_{tt'}$, being equal to: $n_{ft}\,e^{-\f{i}{2}\theta_{fv}\sigma_z}\,n_{ft'}\mone$, does not depend on the Euler angles $\gamma$. The angles $\theta_{fv}$ have to take care of the relations between adjacent $n_{ft}$ and $n_{ft'}$ for any local choice of these Euler angles.

The enlarging of the configuration space thus leads to an additional $\U(1)$ gauge symmetry, acting at each pair $(f,t)$, on the right of $n_{ft}$. This means that each pair $(f,t)$ is equipped with a $\U(1)$ reference frame which can be arbitrarily transformed by $\U(1)$ right multiplication. The angles $\theta_{fv}$ appear as a sort of connection for this gauge symmetry. Let us note that to take care of the $\SU(2)$ gauge invariance, one only has to look at the left of $n_{ft}$. Indeed, these symmetries act on the new variables as:
\begin{align} \label{u1 gauge transfo}
&(k,\lambda)\,\triangleright n_{ft} = k_t\,n_{ft}\,e^{\f{i}{2}\lambda_{ft}\sigma_z} \\
&(k,\lambda)\,\triangleright \theta_{fv} = \theta_{fv} +\eps_{tt'}^f\l(\lambda_{ft} - \lambda_{ft'}\r)
\end{align}
where $\eps_{tt'}^f=\pm1$. Since $t$ and $t'$ are adjacent, the corresponding dual edges share the vertex $v$ along the boundary of the dual face $f$. $\eps_{tt'}^f$ is positive when the path $(t\arr t')$ through $v$ is oriented like $f$, and else negative. The transformations of the angles $\theta_{fv}$ are clearly a discretisation of the usual $\u(1)$ gauge transformations for a connection. They transform so that the holonomies $g_{tt'}$ are indeed independent of the local choices of Euler angles $\gamma$.

Consequently, we will be interested, to make the link with Regge calculus, in $\SU(2)$ and $\U(1)$ gauge invariant quantities. When solving the simplicity constraints, we will show that such angles have the meaning of dihedral angles between tetrahedra $t$ and $t'$, and that the $\U(1)$ subgroup preserving $b_{ft}$, generated by $n_{ft}\,\exp(-i\theta\sigma_z/2)\,n_{ft}\mone$, acquires a real physical meaning.

To get a hint about the meaning of these angles, note that the content of the delta functions in \eqref{gluing measure} can now be used to rewrite the action as a function of $n_{ft}$ and $\theta_{fv}$, in addition to $A_f$ and $g_{vt}$. The relation \eqref{gluing n} can first be inverted to express $g_{tt'}$ in terms of $n_{ft},\ n_{ft'}$ and $\theta_{fv}$. Thus, it corresponds to (partially) solving the discrete analog of the e.o.m. $d_AB=0$ \eqref{daB} for the connection, in terms of the discrete $B$, with a free parameter, $\theta_{fv}$. The holonomy around each face is then given by:
\beq
g_f(t) = n_{ft}\,e^{-\f{i}{2}\theta_f \sigma_z}\,n_{ft}\mone = e^{-\f{1}{A_f}\theta_fb_{ft}}
\ee
where $\theta_f = \sum_{v\in\pp f} \theta_{fv}$. Thus, $\theta_f$ represents the class angle of the rotation $g_f(t)$, and its direction is directly given by that of $b_{ft}$, $\hat{b}_{ft}$. In particular $b_{ft}$ is left invariant by $g_f(t)$ as expected. We can now rewrite the action with the additional variables:
\beq
S_{\mathrm{BF}}\bigl(A_f,n_{ft},\theta_{fv},g_{vt}\bigr) = \sum_f A_f\,\sin\l(\f{\theta_f}{2}\r)
\ee
It is trivially $\SU(2)$ gauge invariant, and also invariant under the $\U(1)$ transformations \eqref{u1 gauge transfo}. Provided that the angles $\theta_{fv}$ are interpreted as dihedral angles and $A_f$ as the area of $f$, it corresponds to the so-called compactified Regge action, proposed in \cite{caselle} and \cite{kawamoto}. To make this statement precise, we need to study the imposition of the simplicity constraints which ensure metricity.

\section{Simplicity and area-angles Regge calculus} \label{area-angle RC}

\subsection{Parametrisation of the constraints}

In the continuum, gravity is described with Spin(4) BF theory when the field $B$ is restricted to a configuration space satisfying:
\beq
\eps_{IJKL}\,B^{IJ}_{\mu\nu}\,B^{KL}_{\lambda\sigma} \propto \eps_{\mu\nu\lambda\sigma}
\ee
where $\mu,\nu,\lambda,\sigma$ are spacetime indices. When the proportionality coefficient is non-vanishing, the theory consists in two branches up to global signs \cite{simplicity}: $B^{IJ} = \pm(e^I\wedge e^J)$ or $(\star B)^{IJ} = \pm(e^I\wedge e^J)$ for a non-degenerate cotetrad 1-form $e$. Gravity corresponds to the second case, up to a global sign in front of the action, whereas the first case will be called to be non-geometric. Its e.o.m. are just $d_A e = 0$, which constrain the torsion to vanish but leave curvature free.

At the discrete level, the constraints only involve the so-called bivectors $B_f^{IJ}(t)$, which can be written in terms of their self-dual and anti-self-dual components, $B_f(t)= b_{+f}(t)\oplus b_{-f}(t)$ due to the splitting $\spin(4) = \su(2)\times\su(2)$. We define the Hodge operator $\star$ to change the sign of the anti-self-dual sector by the following action on $\spin(4)$ elements: $(\star B)^{IJ} = \f{1}{2}\eps^{IJ}_{\phantom{IJ}KL}\,B^{KL}$. The constraints are split into three types :
\begin{alignat}{4}
&\text{Diagonal simplicity}&&\qquad\qquad\qquad \bigl(\star B_f(t)\bigr)_{IJ}\ B_f^{IJ}(t) = 0 \label{diag}\\
&\text{Cross-simplicity}&&\qquad\qquad\qquad \bigl(\star B_f(t)\bigr)_{IJ}\ B_{f'}^{IJ}(t) = 0 \label{off-diag}\\
&\text{Closure relation}&&\qquad\qquad\qquad \sum_{f\subset \pp t} \eps_{ft}\,B_f(t) = 0 \label{closure}
\end{alignat}
where $\eps_{ft}=\pm1$ according to the relative orientation of the dual face $f$ and the dual edge $t$. In the second line, $f$ and $f'$ are triangles of a given tetrahedron. The first constraint, \eqref{diag}, often called diagonal simplicity, expresses the fact that $B_f$ is simple, i.e. it is the anti-symmetrized product of two vectors $B_f^{IJ} = u^{[I}v^{J]}$ (and the same for its Hodge dual). The second constraint, called cross-simplicity constraint, asks for the sum of two bivectors of a tetrahedron to be also simple. Finally, the constraint \eqref{closure} imply that all tetrahedra all closed and is thus mentioned as the closure relation. When these constraints are satisfied for a tetrahedron, its edges can be labelled by 4-vectors only spanning a three-dimensional subspace of $\mathbb{R}^4$. As in the continuum, the bivectors or their Hodge dual are wedge (anti-symmetrized) products of these vectors up to a sign\footnotemark $\eps=\pm1$. To distinguish between the two solutions, \eqref{off-diag} can be changed with:
\beq
\eps_{IJKL}\,B_f^{IJ}(t)\,N_t^K = 0, \qquad\text{or}\qquad B_f^{IJ}(t)\,N_{tI} =0 \label{cross-diag}
\ee
respectively for the geometric and non-geometric sectors. $N_t$ is in both cases interpreted as a unit vector perpendicular to the edges of the tetrahedron $t$. The interesting point is that, in a 4-simplex, simplicity of the five tetrahedra, together with the relations of parallel transport between adjacent tetrahedra
\beq \label{transport bivectors}
B_f(t) = G_{tt'}\,B_f(t')\,G_{tt'}\mone,
\ee
with $G_{tt'} = G_{vt}\mone G_{vt'}$, induce metricity of the 4-simplex. This is due to the fact that two triangles sharing an edge in 4-simplex can always be viewed to be in the boundary of a tetrahedron (this point will be crucial in the present work). Equivalently, one can define the bivectors in the frames of the 4-simplices to be: $B_{ft}(v) \equiv \eps_{ft}\,G_{vt}\,B_{f}(t)\,G_{vt}\mone$, and use the theorems of \cite{bc} within each 4-simplex and of \cite{conrady2} to reglue them. Notice that the sign $\eps$ is then a global sign.

\footnotetext{To eliminate this sign ambiguity, one should ask for a condition on the positivity of tetrahedron volumes, as in \cite{bc}. We will not deal with such a constraint, as it is not present in the known spin foam models for quantum gravity.}

Diagonal and cross-simplicity are easily solved by introducing bivectors $B_{ft}=b_{+ft}\oplus b_{-ft}$ for each tetrahedron sharing a given triangle $f$, and working with their self-dual and anti-self-dual components, parametrized as in the previous section \eqref{b parametrisation}, $b_{\pm ft} =\f{i}{2}A_{\pm f}\,n_{\pm ft}\sigma_z n_{\pm ft}\mone$. The areas $A_{\pm f}$ do not depend on tetrahedron frames since the Spin(4) parallel transport does not affect the norms. Then, \eqref{diag} and \eqref{off-diag} can be solved independently within each tetrahedron, implying relations between the self-dual and anti-self-dual areas $A_{\pm f}$ and directions $n_{\pm ft}$. Indeed, \eqref{diag} means that $b_{\pm ft}$ have equal squared norms, $A_{-f}^2 = A_{+f}^2$. Equivalently, they are related by a rotation $h_{ft}\in\SU(2)$:
\beq \label{diag sol}
\exists\, h_{ft}\in\SU(2)\quad b_{-ft} = \pm h_{ft}\mone\,b_{+ft}\,h_{ft}
\ee
As shown in \cite{etera simone}, in the geometric (respectively non-geometric) sector determined by the sign '-' (resp. '+'), $h_{ft}$ has a clear geometric meaning. The Spin(4) element $H_{ft}=(h_{ft},\mathrm{id})$ maps the reference vector $N^{(0)}=(1,0,0,0)$ to a unit vector $N_{ft}$ orthogonal to $f$ in the sense $(\star B_{ft})^{IJ} N_{ft I} = 0$ in the frame of $t$ (resp. $B_{ft}^{IJ} N_{ft I} = 0$).

$h_{ft}$ thus stands for the choice of a normal to $f$. Cross-simplicity then imposes that such a choice can be done at the level of each tetrahedron, for its four triangles alltogether \cite{fk},\cite{lagrangian BC}:
\beq
b_{-ft} = \pm h_{t}\mone\,b_{+ft}\,h_{t}, \label{cross-simp b}
\ee
$h_t$ representing a unit vector $N_t^I=(h_t,\mathrm{id})^{IJ}N^{(0)}_J$ satisfying the constraint \eqref{cross-diag}. A gauge transformation $K$ acts independently on the self-dual and anti-self-dual sectors according to \eqref{gauge transfo}. It acts on $h_t$ so as to preserve the Spin(4) covariance while imposing the constraints (which relate the self-dual and anti-self-dual sectors) :
\beq \label{transport h}
K\,\triangleright h_t= k_{+t}\,h_t\,k_{-t}\mone
\ee
Due to this specific transformation, $h_t$ can always be gauged-fixed to the identity, which corresponds to the usual time gauge in loop quantum gravity. In this gauge, all tetrahedra are orthogonal to the reference vector $N^{(0)}$. This does {\it not} imply trivial correlations, since quantities must be compared in a common frame. For instance, comparing the normals of two adjacent tetrahedra $t$ and $t'$ can be done by transporting $h_{t'}$ in the frame of $t$, i.e. $g_{+tt'}\,h_{t'}\,g_{-tt'}\mone$. After the gauge-fixing, correlations are encoded into $g_{\pm tt'}$.

Relations \eqref{diag sol} and \eqref{cross-simp b} imply that $h_{ft}$ and $h_t$ differ by an element of the $\U(1)$ subgroup leaving $b_{+ft}$ invariant. Indeed, a triangle in 4d admits an orthogonal plane on which $h_{ft}$ and $h_t$ represent different choices of unit vectors. This orthogonal plane is generated from a normal of reference by action of the $\U(1)$ subgroup preserving $b_{+ft}$. Consequently, as a triangle is shared by several tetrahedra, the vectors $N_t$ are not independent of each other. For two adjacent tetrahedra $t$ and $t'$, $N_t$ and $N_{t'}$ both lie on the plane orthogonal to $\star B_{ft}$ (in the geometric sector), up to parallel transport, provided regluing conditions hold so that the bivectors $B_{ft}$ and $B_{ft'}$ are related following \eqref{transport bivectors}. This geometric information is recorded through an angle $\varphi_{fv} \in[0,4\pi)$ and the relation:
\beq \label{compare normals}
h_t\ \Bigl(g_{+tt'}\,h_{t'}\,g_{-tt'}\mone\Bigr)\mone = n_{ft}\, e^{\f{i}{2}\varphi_{fv}\sigma_z}\, n_{ft}\mone
\ee

Having introduced the normals $N_t$, the constraints can be formulated so as to be linear in the bivectors \eqref{cross-diag},\eqref{cross-simp b}. Without an explicit regluing such that presented in the previous section, equation \eqref{cross-simp b} sets up the framework of the Barrett-Crane spin foam model. The linearity of the action and the constraints in the bivectors together with the formulation of the normals $N_t$ as group elements make the computation of the partition function easy to perform. The regluing is implicitly (and incorrectly) performed via boundary variables for the discretised connection (see details in \cite{lagrangian BC}).

Before gathering the gluing relations of the previous section with the simplicity constraints, we need to rewrite the latter in terms of the area and direction variables. We have self-dual and anti-self-dual areas satisfying $A_{+f}=\pm A_{-f}$, self-dual and anti-self-dual rotations $n_{\pm ft}$ encoding the directions of $b_{\pm ft}$, which are related by the group elements $h_t$. Taking into account the $\U(1)\times\U(1)$ invariance in the definition of $\hat{B}_{ft} = (n_{+ft},n_{-ft})$ :
\beq \label{constrained measure}
\prod_f d\mu\bigl(A_{+f}\bigr) d\mu\bigl(A_{-f}\bigr)\,\delta\bigl(A_{+f} \mp A_{-f}\bigr)\ \prod_{(f,t)} dn_{+ft}dn_{-ft}\ \int \prod_t dh_t\ \prod_{(f,t)} \int_0^{4\pi} d\psi_{ft}\,\delta\Bigl(n_{-ft}\mone\,h_t\mone\,n_{+ft}\,e^{\f{i}{2}\psi_{ft}\sigma_z}\Bigr)
\ee
Diagonal simplicity and the choice of the sector are contained in the one-dimensional delta functions on the areas. Cross-simplicity is imposed by the last $\SU(2)$ delta functions in \eqref{constrained measure}, and $dh_t$ is the $\SU(2)$ Haar measure. The closure relation has to be added. Due to cross-simplicity, only its self-dual components are needed :
\beq \label{closure constraint}
\prod_t \delta\Bigl(\sum_{f\in\pp t} \eps_{ft}\,A_{+f}\,n_{+ft}\sigma_z n_{+ft}\mone\Bigr)
\ee
Taking two copies of \eqref{gluing measure} the regluing relations imposing \eqref{transport bivectors} read :
\beq \label{gluing measure spin4}
\prod_{(f,v)} \int_0^{4\pi} d\theta^+_{fv}\,\delta\Bigl(n_{+ft}\mone\,g_{+tt'}\,n_{+ft'}\,e^{\f{i}{2}\theta^+_{fv}\sigma_z}\Bigr)\ \int_0^{4\pi} d\theta^-_{fv}\,\delta\Bigl(n_{-ft}\mone\,g_{-tt'}\,n_{-ft'}\,e^{\f{i}{2}\theta^-_{fv}\sigma_z}\Bigr)
\ee
They obviously imply that $B_{ft}$ is left invariant by the holonomy based at $t$, $G_f(t)\,B_{ft}\,G_f\mone(t) = B_{ft}$. As a consequence, the normal vector $N_t$ is not preserved by $G_f(t)$ but send to another which is still orthogonal to $\star B_{ft}$:
\beq
\bigl(\star B_{ft}\bigr)^I_{\phantom{I}J}\,\bigl(G_f(t)^J_{\phantom{J}K}\,N_t^K\bigr) = 0
\ee

\subsection{Geometry and relation to area-angle Regge calculus}

After the simplicity constraints are taken into account, the initial variables $B_{ft}$ and $G_{vt}=(g_{+vt},g_{-vt})$ contain all the geometric information. It is then interesting to know how the transition to a certain form of Regge calculus can be explicitly made. We may suspect the angles $\theta^{\pm}_{fv}$ and $\psi_{ft}$ introduced in the measure to be related to dihedral angles. Moreover, having partially solved for the discretised connection with the gluing relations, we expect to switch from the first order setting of BF theory to a second order formalism, provided 4-simplices are not degenerate.

The variables of area-angle Regge calculus are the triangle areas and the 3d dihedral angles $\phi_{ff'}^t$, i.e. the angles between two triangles $f$, $f'$ within the tetrahedron $t$. Areas and 3d angles are naturally subject to constraints, which are going to be discussed below. The 4d dihedral angles $\theta_{tt'}$, measuring the angles between tetrahedra and which appear in the Regge action, are in \cite{area-angleRC} functions of the 3d angles. Using the geometric meaning of the variables $B_{ft}$, $h_t$ and $G_{vt}$, we are going to define some quantities which will naturally be interpreted as areas, 3d and 4d dihedral angles. We will then check that the gluing relations together with simplicity lead to the standard expected relations between these quantities in area-angle Regge calculus. We emphasize that the relations derived in this section are not new complicated relations to be taken care of the path integral, but are simply consequences of the parallel transport conditions \eqref{gluing measure spin4} and of the simplicity constraints as written in \eqref{constrained measure}.

Like in the previous section, instead of the true directions $\hat{b}_{\pm ft}$ of the bivectors, which are in $S^2\times S^2$, we consider the configuration variables as being $(n_{+ft},n_{-ft})\in$ Spin(4), together with the angles $\theta^\pm_{fv}$ and $\psi_{ft}$ which are integrated in \eqref{constrained measure} and \eqref{gluing measure spin4}. This enlarging of the configuration space enables to solve the parallel transport relations for the holonomies and to express cross-simplicity only in terms of group elements. That was already the content of the delta functions in the above measures. Given an assignment of $G_{vt}$ and $(n_{+ft},n_{-ft})$, the regluing and the simplicity constraints ask for the existence of angles such that :
\beq \label{gluing and simplicity}
g_{+tt'} = n_{+ft}\,e^{-\f{i}{2}\theta_{fv}^+\sigma_z}\,n_{+ft'}\mone,\qquad g_{-tt'} = n_{-ft}\,e^{-\f{i}{2}\theta_{fv}^-\sigma_z}\,n_{-ft'}\mone\quad\text{and}\quad n_{-ft} = h_t\mone\,n_{+ft}\ e^{\f{i}{2}\psi_{ft}\sigma_z}
\ee
Let us emphasize that $n_{\pm ft}$ and the angles do not catch all the information about the discrete holonomies, since $G_{tt'}$ is defined as $G_{tt'} = G_{vt}\mone\,G_{vt'}$. Moreover, the group element $G_{vt}$ is the same for the four dual faces sharing the dual edge $t$, while $n_{\pm ft}$ and the angles depend on those faces.

Like in pure BF theory, the Euler angles $\gamma$ of $n_{\pm ft}$ are unphysical (for $n=e^{\f{i}{2}\alpha\sigma_z}e^{\f{i}{2}\beta\sigma_y}e^{\f{i}{2}\gamma\sigma_z}$). The crucial point is that $g_{\pm tt'}$ and $h_t$, as expressed in \eqref{gluing and simplicity}, must not depend on local choices of those Euler angles. The enlarging of the configuration space is thus compensated by an additional $\U(1)\times\U(1)$ gauge symmetry, very similar to \eqref{u1 gauge transfo}:
\begin{align} \label{u1u1 gauge transfo}
&(K,\Lambda)\,\triangleright n_{\pm ft} = k_{\pm t}\,n_{\pm ft}\,e^{\f{i}{2}\lambda_{ft}^\pm\sigma_z} \\
&(K,\Lambda)\,\triangleright \theta_{fv}^\pm = \theta_{fv}^\pm +\eps_{tt'}^f\bigl(\lambda_{ft}^\pm - \lambda_{ft'}^\pm\bigr) \\
&(K,\Lambda)\,\triangleright \psi_{ft} = \psi_{ft}-\bigl(\lambda^+_{ft}-\lambda^-_{ft}\bigr)
\end{align}
The angles $\theta_{fv}^+$ and $\theta^-_{fv}$ form a sort of discrete $\u(1)\oplus\u(1)$ connection. The angle $\psi_{ft}$ transforms so as to preserve the $\U(1)\times\U(1)$ covariance while solving cross-simplicity. Its role is similar to that played by the rotation $h_t$ for the $\SU(2)\times\SU(2)$ gauge symmetry, \eqref{transport h}. In particular, we may choose a gauge fixing such that $h_t=\id$ and $\psi_{ft}=0$, so that the cross-simplicity constraint, i.e. the third equation of \eqref{gluing and simplicity}, reduces to $\delta(n_{-ft}\mone n_{+ft})$. We now consider the geometric quantities of interest, which are both Spin(4) and $\U(1)\times\U(1)$ invariant.

The simplest local observable is the area of a triangle. Since in the continuum the field $B$ is built from the cotetrad, in the lattice the norm of squared $B_{ft}$ represents the squared element of the metric integrated in the directions of the triangle, i.e. its area. Because of diagonal simplicity, $A_{+f}^2=A_{-f}^2$, the self-dual part is sufficient to recover the area:
\beq
\Tr\bigl(B_{ft}^2\bigr) = 2\,\tr\bigl(b_{+ft}\bigr) = -A_{+f}^2
\ee

Considering the products of bivectors within a single tetrahedron then enables to reconstruct its geometry. Indeed, the angle $\phi_{ff'}^t$ between $f$ and $f'$ in $t$ is related to the dot product of the 4-vectors $N_{tJ}\,B_{ft}^{IJ}$ which are orthogonal to triangles in the 3d subspace spanned by $t$. It is easy to see that those dot products correspond to looking at the traces: $\Tr B_{ft} B_{f't}$ \cite{epr}. Because of cross-simplicity, the self-dual components are sufficient. We precisely consider:
\begin{align} \label{3d angle}
\cos \phi_{ff'}^t &= -\eps_{ff'}\ \hat{b}_{+ft}\cdot\hat{b}_{+f't} \\
&= -\eps_{ff'}\ D^{(1)}_{00}\Bigl(n_{+ft}\mone\,n_{+f't}\Bigr)
\end{align}
where $\eps_{ff'}=\eps_{ft}\eps_{f't}$ is the relative orientation of the dual faces $f$ and $f'$. The last equality is a simple computation involving two spins 1/2 which couple only in the spin 1 representation, $\hat{b}\cdot\hat{b}' = \f{1}{2} \tr (n\sigma_z n\mone\, n' \sigma_z n'^{-1}) = D^{(1)}_{00}(n\mone n')$, where $D^{(1)}_{00}$ is the diagonal matrix element in the spin 1 representation for the state $\lvert 1,0\ket$. Using the Euler angles, $g=e^{-\f{i}{2}\alpha\sigma_z} e^{-\f{i}{2}\beta\sigma_y} e^{-\f{i}{2}\gamma\sigma_z}$, it is a very simple function which is in fact the Legendre polynomials $P_1$: $\bra 1,0\rv g \lv 1,0\ket = P_1(\cos\beta) = \cos\beta$. As expected, the $\U(1)$ action on the right of each $n_{+ft}$ is absorbed by the state $\lv1,0\ket$. The 3d angle $\phi_{ff'}^t$ is thus a $\SU(2)$ and $\U(1)$ gauge invariant quantity which is encoded in the correlations between the group elements $n_{+ft}$ and $n_{+f't}$.

As for the internal (4d) dihedral angles between two adjacent tetrahedra, they are defined by the product of the normal vectors, $-N_t\cdot N_{t'}$, up to parallel transport: 
\begin{align}
\cos\theta_{tt'} &= -N_t(v)\cdot N_{t'}(v) \\
&= -\f{1}{2}\,\tr\ h_t\,g_{-tt'}\,h_{t'}\mone\,g_{+tt'}\mone
\end{align}
The last equality simply follows from the standard action of Spin(4) on $\R^4$. The key quantity to be studied is thus the group element $H_{tt'}$, defined in the frame of $t$, which compares the two normals:
\beq \label{def 4d angles}
H_{tt'} = h_t\,g_{-tt'}\,h_{t'}\mone\,g_{+tt'}\mone
\ee
As we have already seen, it can be expressed as an element of the $\U(1)$ subgroup which leaves $b_{+ft}$ invariant \eqref{compare normals}, since $N_t$ and $N_{t'}$ are both in the plane orthogonal to $\star B_{ft}$. We are thus mainly interested in the class angle. Using \eqref{gluing and simplicity}, it can be expressed in terms of the angles $\theta_{fv}^\pm$ and $\psi_{ft}$:
\beq \label{4d angles}
H_{tt'} = n_{+ft}\,e^{\f{i}{2}\bigl(\eps_{tt'}^f(\theta^+_{fv}-\theta^-_{fv})+\psi_{ft}-\psi_{ft'}\bigr)\sigma_z}\,n_{+ft}\mone
\ee
where $\eps_{tt'}^f=\pm1$ is positive when the path $(t\arr t')$ through $v$ is oriented like $f$, and else negative. Let us define the internal dihedral angle $\theta_{tt'}$, whose label is equivalent to $\theta_{fv}$, via:
\begin{align}
\cos\theta_{tt'} &= -\cos\f{1}{2} \Bigl[\theta_{fv}^+-\theta_{fv}^-+\eps_{tt'}^f \bigl(\psi_{ft}-\psi_{ft'}\bigr)\Bigr] \\
\sin\theta_{tt'} & = \eps\,\eps_{tt'}^f\,\sin\f{1}{2} \Bigl[\theta_{fv}^+-\theta_{fv}^-+\eps_{tt'}^f \bigl(\psi_{ft}-\psi_{ft'}\bigr)\Bigr]
\end{align}
where $\eps$ is precisely the global sign ambiguity appearing in the expression of the bivectors in terms of the edge vectors \cite{barrett fairbairn}. It is then convenient to express $H_{tt'}$ with the direction $\hat{b}_{+ft}$ and the dihedral angle between $t$ and $t'$:
\beq \label{4d angles 2}
H_{tt'} = - \cos\bigl(\theta_{tt'}\bigr) + i\eps\,\sin\bigl(\theta_{tt'}\bigr)\, \hat{b}_{+ft}\cdot\vec{\sigma}
\ee
with $\hat{b}_{+ft}\cdot\vec{\sigma} = n_{+ft}\sigma_z n_{+ft}\mone$. 

We have started with lattice BF gauge theory, which is a first order formalism, in which the connection is independent of the field $B$. In the continuum, when simplicity is implemented, the equations of motion $d_A B =0$ reduces to $d_A e=0$ if $e$ is a non-degenerate cotetrad. This latter equation then admits a unique solution $A(e)$, which is said to be compatible with the cotetrad. Here, the gluing conditions have partially solved a discrete analog of the e.o.m. $d_AB = 0$ for the discrete connection, $g_{\pm tt'} = n_{\pm ft}\,e^{-\f{i}{2}\theta^\pm_{fv}\sigma_z}\,n_{\pm ft'}\mone$. Barrett \cite{barrett first order} has emphasized that first order Regge calculus can be defined provided the angles $\theta_{tt'}$ are restricted to be dihedral angles of any well-defined geometry of a 4-simplex. Since here the only geometry is that given by the bivectors, the natural question for our setting is to know if the simplicity constraints together with the gluing relations are sufficient to relate the dihedral angles $\theta_{tt'}$ to the geometry descibed by the bivectors $B_{ft}$, i.e. to the 3d angles $\phi_{ff'}^t$, according to \eqref{relation 3d-4d angles}. This is indeed the case.

\begin{figure} \begin{center}
\includegraphics[width=3cm]{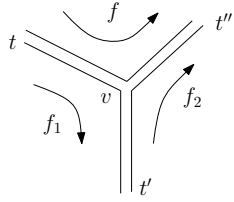}
\caption{ \label{setting} In the dual picture, the three tetrahedra $t$, $t'$ and $t''$ of the 4-simplex $v$ become edges meeting at the vertex $v$. Each triangle being shared by two tetrahedra in $v$, the boundaries of the dual faces in the neibourhood of $v$ are made of two dual edges. For a pair of tetrahedra meeting at $f$, one can equivalently write the dihedral angle $\theta_{tt''}$ between them in terms of the 3d angles using three different intermediate tetrahedra (the boundary of a 4-simplex being made of five tetrahedra). This leads to the constraints \eqref{consistency 3d angles} between the 3d angles, as proposed in \cite{area-angleRC} }
\end{center}
\end{figure}

Consider three tetrahedra $t$, $t'$ and $t''$ in a 4-simplex $v$, each pair meeting at a triangle, according to fig. \ref{setting}. The 4d dihedral angles are defined in a flat 4-simplex in terms of the 3d angles by:
\beq \label{relation 3d-4d angles}
\cos\theta_{tt''} = \f{\cos\phi^{t'}_{f_1f_2}-\cos\phi_{f_1 f}^t\,\cos\phi_{ff_2}^{t''}}{\sin\phi_{f_1 f}^t\,\sin\phi_{ff_2}^{t''}}
\ee
It is important to see that the intermediate tetrahedron $t'$, with triangles $f_1$ and $f_2$, can be changed with another tetrahedron of the 4-simplex. This leads to constraints for area-angle Regge calculus between the 3d angles. The complicated relation \eqref{relation 3d-4d angles} is in fact naturally encoded in the present framework into the constraints written as relations among $\SU(2)$ variables. To obtain it, one can recognize the structure of the multiplication law of $\SU(2)$. The idea is then to express $H_{tt''}$ in terms of $H_{tt'}$ and $H_{t't''}$. From its definition \eqref{def 4d angles}, one simply has:
\beq
H_{tt''} = H_{tt'}\,g_{+tt'}\,H_{t't''}\,g_{+tt'}\mone \label{compo H}
\ee
This is the key identity in our setting. From the definition of $H_{tt'}$, it is quite trivial and mainly due to the fact that parallel tansport between $t$ and $t''$ through their common triangle $f$ can be performed using $t'$ as a pivot, along the boundary of the dual faces $f_1$ and $f_2$ instead: $g_{+tt''} = g_{+vt}\mone\,g_{+vt'}\,g_{+vt'}\mone\,g_{vt''} = g_{+tt'}\,g_{+t't''}$. However, when $H_{tt'}$ is expressed as a $\U(1)$ element preserving $b_{+ft}$, \eqref{4d angles 2}, it becomes a non-trivial statement. Taking the trace of equation \eqref{compo H}, using the regluing conditions along $f_1$ between $t$ and $t'$, one obtains:
\beq
-\cos \theta_{tt'} = \f{1}{2}\tr\Bigl[\bigl(-\cos\theta_{tt'}+i\eps\,\eps_{tt'}^{f_1}\,\sin\theta_{tt'}\,n_{+f_1t'}\sigma_z n_{+f_1t'}\mone\bigr) \bigl(-\cos\theta_{t't''}+i\eps\,\eps_{t't''}^{f_2}\,\sin\theta_{t't''}\,n_{+f_2t'}\sigma_z n_{+f_2t'}\mone\bigr)\Bigr]
\ee
from which it follows that:
\beq
\cos \theta_{tt'} = - \cos\theta_{tt'}\ \cos\theta_{t't''} + \sin\theta_{tt'}\ \sin\theta_{t't''}\ \cos\phi_{f_1f_2}^{t'}
\ee
This relation is naturally $\SU(2)$ and $\U(1)$ invariant. For non-degenerate configurations such that $\sin\theta_{tt'}\neq 0$, it gives the 3d angles as functions of the 4d angles. It can also be inverted, precisely yielding to \eqref{relation 3d-4d angles}. Interestingly, since \eqref{compo H} is an equation between $\SU(2)$ elements, its projection onto $\vec{\sigma}$ also gives relations between the 3d and 4d angles, which are only $\SU(2)$ covariant. However, their geometric meaning is unclear to us.

We have almost completed the proof that \eqref{gluing and simplicity} and the closure relation contain the constraints of area-angle Regge calculus. Indeed, the expression of the angles $\theta_{t_1t_2}$, \eqref{relation 3d-4d angles}, has been established for any intermediate tetrahedron. The 3d angles hence satisfy:
\beq \label{consistency 3d angles}
\f{\cos\phi_{f_1f_2}^t - \cos\phi_{f_1f}^{t_1}\cos\phi_{ff_2}^{t_2}}{\sin\phi_{f_1f}^{t_1}\sin\phi_{ff_2}^{t_2}}
 = \f{\cos\phi_{f'_1f'_2}^{t'} - \cos\phi_{f'_1f}^{t_1}\cos\phi_{ff'_2}^{t_2}}{\sin\phi_{f'_1f}^{t_1}\sin\phi_{ff'_2}^{t_2}}
\ee
which implies a consistent gluing of tetrahedra. Here, $t_1$ and $t_2$ share the triangle $f$, $t_1$ and $t$ share $f_1$, and $t$ and $t_2$ share $f_2$ (and similarly for the primed labels). Moreover, by contracting the closure relation \eqref{closure} with a specific bivector $\eps_{ft}B_{ft}$, we obtain a gauge invariant form:
\beq \label{closure 2}
A_{+f}-\sum_{f'\neq f\subset\pp t}A_{+f'}\,\cos\phi_{ff'}^t = 0
\ee
The constraints \eqref{consistency 3d angles} and \eqref{closure 2} are precisely proposed in \cite{area-angleRC} as a set of constraints for area-angle calculus.

\subsection{The action and the Immirzi parameter}

Before studying how these variables can be useful to build some old and new spin foam models, we need to write an action. It is expected to look like a (compactified) Regge action, but the role of the Immirzi parameter in this context is badly known (in particular, it was not introduced in \cite{area-angleRC}). Spin(4) BF theory can be equivalently formulated with the field $B$ or its Hodge dual. It turns out that even for gravity, when $B=\star(e\wedge e)$, a term proportional to $\tr(\star B\wedge F)$ can be added to the action without modifying the e.o.m. (at least for pure gravity, without matter), the proportionality coefficient being known as the inverse of the Immirzi parameter $\gamma$. Defining $\gamma_\pm = \pm(1\pm\gamma\mone)$, the naive action is thus:
\beq \label{naive action}
I_\gamma^{\mathrm{naive}} = \sum_f \tr\Bigl(\bigl(1+\gamma\mone\star\bigr)B_{ft}\,G_f(t)\Bigr) = \sum_f \gamma_+\tr\bigl(b_{+ft}\,g_{+f}(t)\bigr) - \gamma_-\tr\bigl(b_{-ft}\,g_{-f}(t)\bigr)
\ee
For each triangle, this action needs a base point, i.e. a tetrahedron of reference. The gluing relations between bivectors of the same triangle, taking into account the structure of parallel transport, makes it independent from the choice of the base points. However, we clearly see that simplicity, especially when rewritten as in \eqref{cross-simp b}, leads to traces of $b_{+ft}$ times a linear combination of $\SU(2)$ elements. This unnatural situation can lead to spurious measure factors in the functional integral, as shown in \cite{lagrangian BC}. It is far more natural to use instead the group structure, which corresponds at first order to the sum of the Lie algebra elements. In this context, the Immirzi parameter is introduced as a coefficient for these algebra elements, as already proposed in \cite{lagrangian BC}. Let us define, for $g=\cos\theta/2+i\sin\theta/2\,\hat{n}\cdot\vec{\sigma}\,\in\SU(2)$ of class angle $\theta$ and direction $\hat{n}$, the element of class angle $\alpha\theta$ and same direction:
\beq
g^\alpha = \cos\bigl(\alpha\f{\theta}{2}\bigr) + i\,\sin\bigl(\alpha\f{\theta}{2}\bigr)\,\hat{n}\cdot\vec{\sigma}
\ee
For integral $\alpha$, it is obviously the group multiplication. To introduce the Immirzi parameter, we may propose: $\sum_f \tr(b_{+ft}\,g_{+f}^{(1+\gamma\mone)}(t)) + \tr(b_{-ft}\,g_{-f}^{(1-\gamma\mone)}(t))$. The cross-simplicity constraints \eqref{cross-simp b}, $b_{-ft}=-h_t\mone b_{+ft} h_t$, still lead to sums of group elements: $\tr\,b_{+ft} (g_{+f}^{\gamma_+}(t)-h_t g_{-f}^{-\gamma_-}(t) h_t\mone)$. We thus change that expression by using the group multiplication which corresponds at first order to the sum of Lie algebra elements like in the continuum action.
\beq \label{action 1}
I_\gamma\bigl(b_{+ft},h_t,g_{\pm vt}\bigr) = \sum_f \tr\bigl(b_{+ft}\,g_{+f}^{\gamma_+}(t)\,h_t\,g_{-f}^{\gamma_-}(t)\,h_t\mone\bigr)
\ee
The regluing relations and the simplicity constraints ensure that it is gauge invariant and independent from the choice of the tetrahedron of reference for each triangle. Note that \eqref{action 1} uses cross-simplicity for each tetrahedron of reference. The sign '-' in front of $(1-\gamma\mone)$ for the anti-self-dual $g_{-f}^{-(1-\gamma\mone)}$ is due to the sign '-' in the constraints for the geometric case, $b_{-ft} = -h_t\mone\,b_{+ft}\,h_t$. Its removing corresponds to the choice $b_{-ft} = h_t\mone\,b_{+ft}\,h_t$.

Another proposal is to impose cross-simplicity for each tetrahedron of reference via the group multiplication, like in \eqref{action 1}, but to keep separately the contribution of the Immirzi parameter:
\beq \label{action 2}
\tl{I}_\gamma\bigl(b_{+ft},h_t,g_{\pm vt}\bigr) = \sum_f \tr\bigl(b_{+ft}\,g_{+f}(t)\,h_t\,g_{-f}\mone(t)\,h_t\mone\bigr) + \gamma\mone\tr\bigl(b_{+ft}\,g_{+f}(t)\,h_t\,g_{-f}(t)\,h_t\mone\bigr)
\ee
It is also $\SU(2)$ gauge invariant and independent from the choice of the base points.

Notice that the naive continuum limits of $I_\gamma$ and $\tl{I}_\gamma$ are exactly the same as that of the naive action $I_\gamma^{\mathrm{naive}}$, \eqref{naive action}. Consider a chart where the typical length of edges of $f$ is of order $\varepsilon$. When $\varepsilon$ goes to zero, we can use the expansion: $g_{\pm f}\approx 1+\varepsilon^2 F_{\pm \vert f}$, where $F_{\pm \vert f}$ is the component of the curvature along the directions of the face dual to the triangle. This reduces $I^{\mathrm{naive}}_\gamma$ and $\tl{I}_\gamma$ to the same expression: $\sum_f \gamma_+\varepsilon^2\tr (b_{+ft}F_{+\vert f}(t)) - \gamma_-\varepsilon^2\tr (b_{-ft}F_{-\vert f}(t))$. Also, with the Immirzi parameter, $g_{\pm f}^{\gamma_\pm}\approx 1+\gamma_\pm\varepsilon^2 F_{\pm \vert f}$. Inserting this expansion into \eqref{action 1}, we are lead to the same result to first order in $\varepsilon^2$, and thus to the same continuum limit.

If we again promote the angles $\theta_{fv}^{\pm}$ and $\psi_{ft}$ appearing in the measure \eqref{gluing measure spin4}, \eqref{constrained measure} to configuration variables, we can easily express the action $I_\gamma$ with only areas and angles. Indeed, from \eqref{gluing and simplicity}, the holonomies around triangles are determined by the directions of the bivectors and the self-dual and anti-self-dual deficit angles $\theta^\pm_f = \sum_{v\in\pp f}\theta^\pm_{fv}$:
\beq
g_{+f}(t) = n_{+ft}\,e^{-\f{i}{2}\theta_f^+\sigma_z}\,n_{+ft}\mone\qquad\text{and}\qquad 
g_{-f}(t) = n_{-ft}\,e^{-\f{i}{2}\theta_f^-\sigma_z}\,n_{-ft}\mone
\ee
The angles of the rotations $g_{\pm f}^{\gamma_\pm}$ are simply $\gamma_\pm\theta^\pm_f$.
This leads to the following simple forms:
\begin{align} \label{grav action}
I_\gamma\bigl(A_{+f},k_{+ft},\theta^\pm_{fv},h_t,\psi_{ft},g_{\pm vt}\bigr) &= \sum_f A_{+f}\,\sin\Biggl(\f{\gamma_+}{2}\,\theta^+_f + \f{\gamma_-}{2}\,\theta^-_f\Biggr) \\
 &= \sum_f A_{+f}\,\sin\Biggl(\f{1}{2}\bigl(\theta^+_f-\theta^-_f\bigr) + \f{1}{2\gamma}\bigl(\theta^+_f+\theta^-_f\bigr)\Biggr)
\end{align}
and for $\tl{I}_\gamma$:
\beq \label{grav action 2}
\tl{I}_\gamma\bigl(A_{+f},k_{+ft},\theta^\pm_{fv},h_t,\psi_{ft},g_{\pm vt}\bigr) = \sum_f A_{+f}\,\sin\f{1}{2}\bigl(\theta^+_f-\theta^-_f\bigr) + \gamma\mone\,A_{+f}\,\sin\f{1}{2}\bigl(\theta^+_f+\theta^-_f\bigr)
\ee
As expected, $A_{+f}$ represents the area of the triangle $f$ as a function of the edge lengths once simplicity is taken into account. We recognize the usual, geometric, deficit angles, $\theta_f=(\theta^+_f-\theta^-_f)/2$. Notice that $\theta_f$ is {\it in fine} insensitive to the angles $\psi_{ft}$ appearing in the simplicity constraints \eqref{gluing and simplicity} and which play a role in the dihedral angles \eqref{4d angles}. But we also see the appearance of some non-geometric deficit angles, given by the sum of the self-dual and anti-self-dual angles, instead of their difference, due to the presence of the Immirzi parameter.

In the continuum, the e.o.m. $d_AB =0$ with $B=\star(e\wedge e)$ for a non-degenerate cotetrad leads to $d_Ae=0$. In the lattice and for nondegenerate sets of bivectors, simplicity and parallel transport make sure that the edge vectors $e_\ell(t)$ defined in local frames of different tetrahedra are related by parallel transport, up to a change of sign \cite{conrady2}. In particular, the holonomy $G_f$ around a dual plaquette leaves the edges of the corresponding triangle invariant up to signs:
\beq \label{edge transport}
G_f(t)^I_{\phantom{I}J}\,e_\ell(t)^J = e^{i\pi\sum_t\eta_t}\,e_\ell(t)^I
\ee
where $\eta_t=0,1$ and the sum is over the dual edges around the dual face $f$. Again in the continuum and for a nondegenerate cotetrad, the equation $d_Ae=0$ can be solved giving a unique connection $A(e)$ compatible with $e$. This connection is such that the Immirzi term of the action, $e^I\wedge e^J\wedge F_{IJ}(A)$, identically vanishes. We would like our discrete setting to also consistently imply the disappearance of the Immirzi parameter from the e.o.m. (involving constraints between the angles $\theta^+_f$ and $\theta^-_f$) for nondegenerate configurations. The parameter $\gamma$ only appears in the action for the closed holonomies $G_f(t)$, so we do not need to carefully look at each step of parallel transport between adjacent tetrahedra. Without taking \eqref{edge transport} into account, $G_f(t)$ only leaves $B_{ft}$ invariant and contains {\it a priori} terms proportional to $(\theta_f^++\theta^-_f)$ as well as $\theta_f=(\theta_f^+-\theta^-_f)/2$:
\begin{align}
G_f(t) &= e^{\f{-1}{A_{+f}}(\theta^+_f b_{+ft} - \theta^-_f b_{-ft})} \\
&= e^{\f{-1}{2A_{+f}}(\theta_f^++\theta_f^-)\star B_{ft}}\ e^{\f{-1}{A_{+f}}\theta_f B_{ft}}
\end{align}
When simplicity is imposed, the edge geometry and in particular equation \eqref{edge transport} have to be considered. It is clear that the terms proportional $\theta_f$ are not involved, since $B_{ft}^{IJ}\,e_{\ell J}(t)=0$ in the geometric sector. Thus, the non-geometric angle $(\theta^+_f+\theta^-_f)$ is responsible for the phase factor of \eqref{edge transport}:
\begin{align}
&\theta^+_f+\theta^-_f = 2\pi\sum_t\eta_t\ (\mathrm{mod}\,4\pi)& \label{away immirzi} \\
&G_f(t) = e^{-\f{\pi}{A_{+f}}(\sum_t\eta_t)\star B_{ft}}\ e^{\f{-1}{A_{+f}}\theta_f B_{ft}}\qquad &&\text{for non-degenerate configurations.} \label{on-shell G}
\end{align}
When the phase factor is trivial, this is a well-known property of Regge calculus. Considering a plane in $\R^4$ generated by the vectors $e_1$ and $e_2$, only a $\U(1)$ subgroup leaves any vector of the plane invariant, which is that generated by $B_{12}=\star(e_1\wedge e_2)$. The plane is stable under the action of the rotations generated by $\star B_{12}=e_1\wedge e_2$ and \eqref{on-shell G} allows a reflection, $e_{1,2}\arr-e_{1,2}$, i.e. a rotation of angle $\pi$ generated by $\star B_{12}$. The form of the on-shell holonomy \eqref{on-shell G} implies:
\beq \label{link g+-}
h_t\mone\ g_{+f}(t)\,h_t\,g_{-f}(t) = e^{i\pi\sum_t\eta_t}\,\id
\ee
This can be interpreted in the time gauge, $h_t=\id$, in which the edge vectors of $t$ span the 3d space orthogonal to $N^{(0)}$. For a triangle $f$, the holonomy takes the form $G_f(t)=(g_{+f}(t),g_{+f}\mone(t))$, or $G_f(t)=(g_{+f}(t),-g_{+f}\mone(t))$. In the first case, it is a boost, in analogy with the Lorentzian terminology, whose axis is the normal $N_f^t$ to $f$ within the 3d space of $t$, i.e. a rotation generated by $N^{(0)}\wedge N_f^t$. In the second case, it is combined with a reflection of $f$.

Equation \eqref{link g+-} has also interesting consequences in the dual picture in terms of normals to tetrahedra. Given the rules for parallelly transporting $h_t$, \eqref{transport h}, this equation means that the normal vector $N_t$ is left invariant by the group element $\star G_f(t)\equiv (g_{+f}(t),\pm g_{-f}\mone(t))$. A similar relation was obtained in \cite{lagrangian BC}, while dealing with Spin(4) BF theory constrained by diagonal simplicity only, in the non-geometric sector. In this situation, $h_t$ is replaced with $h_f$ which stands for the choice of a normal $N_f=(h_f,\id)N^{(0)}$ to the triangle in the sense $B_f^{IJ}N_{f J}=0$. The equation $h_f\mone(t)\,g_{+f}(t)\,h_f(t)\,g_{-f}(t) = \id$ then results from varying the action with respect to the bivector. In the geometric sector, one has instead $h_f\mone(t)\,g_{+f}(t)\,h_f(t)\,g_{-f}\mone(t) = \id$, meaning that $N_f$ is left invariant by $G_f$. It is thus natural that \eqref{link g+-} entails the disappearance of the Immirzi parameter of the action, when seen as an e.o.m. of the non-geometric sector. Notice indeed that the quantity of the l.h.s. of \eqref{link g+-} is precisely that entering the action $\tl{I}_\gamma$, \eqref{action 2}, with the coefficient $\gamma\mone$. It thus becomes on-shell independent of $\gamma$:
\beq \label{action 2 on-shell}
\tl{I}_\gamma = \sum_f A_{+f}\,\sin\theta_f
\ee
As for the action $I_\gamma$, it looses its dependence on $\gamma$ and reduces to \eqref{action 2 on-shell} when $\gamma\mone$ is an integer.

\section{Spin foams for BF theory and insertions of local observables} \label{SF BF}

In the previous sections, we have discussed an improved setting to discretize gravity from BF theory with constraints. We have proposed to explicitly introduce the parallel transport relations for bivectors in the discretized functional integral. We here check that in the absence of constraints, the standard spin foam model for BF theory is recovered, fixing in addition the integration measure for the area. Moreover, the present framework allows to insert local bivector observables, corresponding to areas and 3d dihedral angles when simplicity is fulfilled, and explicitly compute their spin foam quantisation within $\SU(2)$ BF theory. It is certainly not the most straightforward road to obtain those results, but it introduces the methods to be used in the presence of the simplicity constraints and which then turn out to allow a straightforward derivation the new spin foam models \cite{lagrangian SF}.

\subsection{BF theory}

As we have seen, the field $B$ in $\SU(2)$ BF theory can be seen as a multiplier imposing the vanishing of the curvature. Thus, the integration over $B$ in the partition function projects onto these configurations. At the discrete level, the holonomy around a dual face $f$ is constrained to be $\pm\id$, or equivalently, its projection onto $\SO(3)$ is the identity:
\begin{align} \label{Zbf}
Z_{\mathrm{BF}} &= \int \prod_{(t,v)} dg_{vt}\ \prod_f \delta_{\SO(3)}\bigl( g_f\bigr) \\
 &= \sum_{\{j_f\in\N\}} \int \prod_{(t,v)} dg_{vt}\ \prod_f d_{j_f}\,\chi\bigl(g_f\bigr)
\end{align}
where $dg$ is the $\SU(2)$ Haar measure, and $d_j = (2j+1)$ and $\chi_j(g)$ are respectively the dimension of the representation of spin $j$ and its character. If $g=e^{i\phi\hat{n}\cdot\vec{\sigma}}$ for $\phi\in[0,2\pi)$, then: $\chi_j(g) = \f{\sin d_j\phi}{\sin\phi}$.

Using the variables introduced in the previous sections, the partition function is defined by:
\beq \label{def Zbf}
Z_{\mathrm{BF}} = \int \prod_{(t,v)} dg_{vt}\ \prod_{(f,t)} dn_{ft}\ \prod_{(f,v)} \Bigl[d\theta_{fv}\,\delta\bigl(n_{ft}\mone\,g_{tt'}\,n_{ft'}\,e^{\f{i}{2}\theta_{fv}\sigma_z}\bigr)\Bigr] \prod_f d\mu\bigl(A_f\bigr) e^{iA_f\sin\f{\theta_f}{2}}
\ee
where $dn$ is the $\SU(2)$ Haar measure, $d\theta$ the Haar measure over $\U(1)$ (between 0 and $4\pi$), while $d\mu(A)$ will be chosen so that this expression reproduces the standard result \eqref{Zbf}.

It is clear that for each face the delta functions can be used to eliminate all $n_{ft}$ but one, whose tetrahedron is the base point, say $n_{ft^\star}$, giving: $\delta(n_{ft^\star}\,g_f(t^\star)\,n_{ft^\star}\mone\,e^{\f{i}{2}\theta_f\sigma_z})$. Integrations over the angles $\theta_{fv}$ can be reduced to an integral over a single angle $\theta_f$. In analogy with Regge calculus, it can be said to represent the deficit angle since it is the class angle of the holonomy $g_f(t)$.

The spin foam formalism is a way to write down the partition function as a state-sum whose data come from the representation theory of the considered groups. We thus need to expand the ingredients of $Z_{\mathrm{BF}}$ into $\U(1)$ and $\SU(2)$ harmonic modes. First, for each face, the regluing conditions lead to a delta over the conjugacy class $\theta_f$ for the holonomy:
\beq \label{conj class}
\int_{\SU(2)} dn\ \delta\bigl(n\,g\,n\mone\,e^{\f{i}{2}\theta\sigma_z}\bigr) = \sum_{j\in\f{\N}{2}} \chi_j\bigl(g\bigr)\,\chi_j\bigl(e^{\f{i}{2}\theta\sigma_z}\bigr)
\ee
In particular, $\chi_j(e^{\f{i}{2}\theta\sigma_z}) = \sum_{p=-j}^j e^{-ip\theta}$.

As mentioned in section \ref{Lattice BF gauge theory}, the area can be taken in $\R$ or in $\R_+$. If $A_f$ is integrated on $\R$ with the Lebesgue measure, then $\theta_f$ is constrained to be 0 or $2\pi$. It directly follows that $g_f=\pm\id$ and so \eqref{Zbf}. It is however instructive, and in the spin foam spirit, to use expansions whose data can be interpreted and considered as physically relevant. Keeping in mind the issue of including the simplicity constraints, we consider $A_f\in\R_+$ and expand the exponential of $i$ times the action over $\U(1)$ modes:
\beq \label{exp action bf}
e^{iA\sin\f{\theta}{2}} = \sum_{m\in\Z} J_m(A)\,e^{im\f{\theta}{2}}
\ee
This is the Jacobi-Anger expansion, which defines the Bessel functions $J_m$ of the first kind: $J_m(A)=\int_0^\pi \f{d\theta}{\pi\, i^m}\,\cos(m\theta)\,e^{iA\cos\theta}$. The Bessel functions satisfy some interesting properties \cite{abramo}. Their integral over $\R_+$ is normalized to 1 for $m\in\N$, and important symmetries are: $J_{-m}(A) = J_m(-A) = (-1)^mJ_m(A)$. Combining those results, we have:
\beq \label{int area}
\int_{\R_+}dA\ e^{iA\sin\f{\theta}{2}} = 1 + \sum_{m>0} \bigl(e^{im\f{\theta}{2}} + (-1)^m\,e^{-im\f{\theta}{2}}\bigr)
\ee
Integrating $\theta$ then leads to, for $j\in\N/2$:
\beq
\sum_{m>0}\sum_{p=-j}^j \int_0^{4\pi} \f{d\theta}{4\pi}\ e^{-ip\theta}\bigl(e^{im\f{\theta}{2}} + (-1)^m\,e^{-im\f{\theta}{2}}\bigr) = \sum_{m>0}\sum_{p=-j}^j \delta_{m,2p}\bigl(1+(-1)^m\bigr)
\ee
This expression is vanishing unless $m$ is even, which implies that $j$ must be an integer. If these conditions are fulfilled, we get, for $m=2m'$: $2\sum_{m'=1}^j 1 = 2j$. As for the term $m=0$ of \eqref{exp action bf}, the integration over $\theta$ is trivial and restricts $j$ to be integral. Gathering these results:
\beq \label{Zbf 2}
Z_{\mathrm{BF}} = \int \prod_{(t,v)} dg_{vt}\ \prod_f \sum_{j\in\N} \bigl(2j+1\bigr)\,\chi_j\bigl(g_f\bigr)
\ee
It is finally easy to check that $\sum_{j\in\N} (2j+1)\,\chi_j(g) = \delta_{\SO(3)}(g)$, as desired.

Let us finally write $Z_{\mathrm{BF}}$ as a sum over spin foams, by integrating the group elements $g_{vt}$. Each of them appears four times, once for each face of the tetrahedron $t$, in the representations, say, $j_1,j_2,j_3$ and $j_4$. Such a group averaging can be written as the identity on the invariant space carrying these representations, $\mathrm{Inv}(j_1,j_2,j_3,j_4)$, under the form of a sum over a complete orthogonal basis. The basis elements, called intertwiners, are specified by the choice of a 2 by 2 pairing and of a (internal or virtual) representation between both pairs. More precisely, we define:
\beq
\iota^{j_1 j_2,i,j_3 j_4}_{m_1 m_2 m_3 m_4} = \sum_{m=-i}^i \begin{pmatrix} j_1 & j_2 & i \\ m_1 & m_2 & m \end{pmatrix} (-1)^{i-m} \begin{pmatrix} i & j_3 & j_4 \\ -m & m_3 & m_4 \end{pmatrix}
\ee
which will be shortly denoted $\iota_{m_a}(j_a,i)$, although it does not make the pairing explicitly appear. The chosen pairing is here $j_1$ with $j_2$, and the internal representation $i$. The result of integrating four matrix elements reads:
\beq \label{int g4}
\int_{\SU(2)} dg\ \prod_{a=1}^4 D^{(j_a)}_{m_a n_a}(g) = \sum_i d_i\ \iota_{m_a}(j_a,i)\,\iota_{n_a}(j_a,i)
\ee
Then, when an element $g_{vt}$ is integrated out, it leads to two intertwiners respectively due to the 4-simplex $v$ and tetrahedron $t$ indices. Since a tetrahedron is shared by two 4-simplices $v$ and $v'$, the previous formula applies twice and the tetrahedron is labelled with two intertwiners of representations $i_{tv}$ and $i_{tv'}$ carrying indices from the tetrahedron ends of $g_{vt}$ and $g_{v't}$. Moreover, the group element $g_{vt}$ always appears composed with $g_{v't}\mone$, so that both intertwiners, provided they have the same pairing, are contracted according to the orthogonality relation:
\beq
\sum_{\{m_a\}} \iota_{m_a}(j_a,i)\, \iota_{m_a}(j_a,i') = \f{\delta_{i,i'}}{d_i}
\ee
As for intertwiners with indices from 4-simplices, the five intertwiners corresponding the five tetrahedra meeting at a 4-simplex $v$ are contracted through the ten faces of $v$, and thus according to the structure of a 15j-symbol from $\SU(2)$ recoupling theory \cite{jucys}. Thus, the partition function can be written \cite{ooguri}:
\beq
Z_{\mathrm{BF}} = \sum_{\{j_f\in\N\}} \sum_{\{i_t\}}\ \prod_f d_{j_f}\ \prod_t d_{i_t}\ \prod_v \mathrm{15j}\bigl(j_f,i_t\bigr)
\ee

\subsection{Insertions of areas and 3d angles}

The data appearing in the sum over spin foams are usually interpreted through the canonical quantization \cite{carlo 3d}, \cite{baez BF}. In particular, the representations $j_f$ labelling dual faces stand for lengths of edges in 3d (this is in fact the very definition of the Ponzano-Regge model) and areas of triangles in 4d. Using the loop quantization, the spectrum of these operators can be computed at the quantum level and their square scales as $j_f(j_f+1)$ \cite{carlo primer}. These results can also be obtained from the covariant quantization, by computing quantum average with the functional integral and insertions of states exciting the dual faces. It has been properly done recently in 3d \cite{etera ryan}. We do not propose this treatment here, the issue of the gauge-fixing being more subtle than in 3d, due to the reducibility of the gauge transformations \cite{blau}. But we are rather interested in the translation of insertions of local observables into the language of spin foams.

First consider a given dual face $f^\star$, and insert into $Z_{\mathrm{BF}}$, \eqref{def Zbf}, the local observable $A_{f^\star}^2$:
\beq
Z_{\mathrm{BF}} = \int \prod_{(t,v)} dg_{vt} \prod_{(f,t)} dn_{ft} \prod_{(f,v)} \Bigl[d\theta_{fv}\,\delta\bigl(n_{ft}\mone\,g_{tt'}\,n_{ft'}\,e^{-\f{i}{2}\theta_{fv}\sigma_z}\bigr)\Bigr] \prod_{f\neq f^\star} d\mu\bigl(A_f\bigr) e^{iA_f\sin\f{\theta_f}{2}} \Bigl[d\mu\bigl(A_{f^\star}\bigr)\,A_{f^\star}^2\, e^{iA_{f^\star}\sin\f{\theta_{f^\star}}{2}}\Bigr]
\ee
The result \eqref{Zbf 2} is naturally not modified for $f\neq f^\star$. As previously, the variables $n_{f^\star t}$ are eliminated using the gluing relations, except one which leads to \eqref{conj class} with $g=g_{f^\star}$ and $\theta=\theta_{f^\star}$. The representation $j_{f^\star}$ is also constrained to be integral because of the integral over $\theta$. The change comes from the integral over the area:
\beq
\int_{\R_+} dA\,A^2\,J_m\bigl(A\bigr) = m^2-1,\qquad\text{for}\ m\in2\N
\ee
The important point is that, as before, the sum over the angular momentum $m_{f^\star}$, which is the variable dual to $\theta_{f^\star}$, can be explicitly performed. In addition to the standard dual face weight, $2j+1$, it leads to the insertion into \eqref{Zbf 2} of the following quantum area:
\beq
A_j = \f{4}{3}\Bigl[ j\bigl(j+1\bigr)-\f{3}{4} \Bigr] = \f{1}{3}\bigl(d_j^2 - 4\bigr)
\ee
This spectrum gives a negative value for $j=0$, would vanish for $j=1/2$, and then increases like $\f{4}{3}j(j+1)$. Up to a coefficient $1/3$, this corresponds to the result obtained in \cite{etera ryan} in 3d, with a precise formulation including a gauge fixing process. As explained by the authors, the result crucially depends on measure factors and the way curvature is described in the lattice theory. It is thus an interesting result, given the different setting chosen here, which involves different measures and computations.

Let us come back to the expansion \eqref{exp action bf}. For a given area, all $\U(1)$ modes are involved. But after integration over $A$, we are free to interpret each component $e^{im\theta/2}$ as a Regge action with deficit angle $\theta$ and quantized area $m$. Areas are usually rather quantized as spins of $\SU(2)$ representations. And indeed, in the previous computations, we see that the sum over $m$ can be explicitly performed, leading to standard results. However, in more intricate situations, this sum may not be possible to perform and $m$ may be interpreted as a quantized area. To convince oneself, let us apply to the same expansion to the action $I_\gamma$ \eqref{grav action}:
\beq
\exp\Bigl(iA_f\sin\bigl(\f{\gamma_+}{2}\theta^+_f + \f{\gamma_-}{2}\theta^-_f\bigr)\Bigr) = \sum_{m_f\in\Z} J_{m_f}\bigl(A_f\bigr)\,e^{\f{i}{2}\gamma_+ m_f \theta^+_f} e^{\f{i}{2}\gamma_-m_f\theta^-_f}
\ee
Defining $m^\pm_f = \gamma_\pm m_f$, we obviously have:
\beq
m_f = m_f^-/\gamma_- = m_f^+/\gamma_+
\ee
This is precisely the relation usually written between the spins quantizing the areas, \cite{fk}, \cite{epr}. We would have indeed written this relation if we have used the bivectors $B^{(\gamma)}_{ft} = b^{(\gamma)}_{+ft}\oplus b_{-ft}^{(\gamma)}$, with $b_{\pm ft}^{(\gamma)} = \pm\gamma_\pm b_{\pm ft}$ to define the theory. Diagonal simplicity would have then implied: $A_{-f}/\gamma_- = A_{+f}/\gamma_+$ for the newly defined self-dual and anti-self-dual areas. This strongly suggests that the information concerning the areas are recorded in the spin foam into the $\U(1)$ modes of the Regge action.

\medskip

Let us now study the insertion of the 3d angle, $\cos \phi_{ff'}^t$, which is the angle between the directions of $\hat{b}_{ft}$ and $\hat{b}_{f't}$ within $t$. Consider two triangles $f_1$ and $f_2$ in $t$. This angle can be expressed as: $D_{00}^{(1)}(n_1\mone n_2)$. We naturally expect its translation into spin foams to involve the intertwining representation $i_t$, since $j_1$ and $j_2$ are coupled through $i_t$, if the pairing of the intertwiner is well chosen. We thus write: $D_{00}^{(1)}(n_1\mone\,n_2) = \sum_{A=-1}^1 D^{(1)}_{0A}(n_1\mone)\,D^{(1)}_{A0}(n_2)$. After integration over areas and dihedral angles, the weight for $f_1$ is:
\beq \label{insertion 3d angle}
\Biggl\{\sum_{j_1\in\N} d_{j_1} D^{(j_1)}_{00}(n_1\mone g_1n_1) + \sum_{m_1>0}\sum_{j_1\in\f{\N}{2}} d_{j_1}\Bigl( D^{(j_1)}_{\f{m_1}{2}\f{m_1}{2}}(n_1\mone g_1n_1) + (-1)^{m_1} D^{(j_1)}_{-\f{m_1}{2}-\f{m_1}{2}}(n_1\mone g_1n_1)\Bigr)\Biggr\}\,D^{(1)}_{0A}(n_1\mone)
\ee
As before, though it has not been explicitly precised, since $D^{(j)}_{-m-m}(g)=\overline{D^{(j)}_{mm}(g)}=D^{(j)}_{mm}(g\mone)$, the partition function sums over the orientation of each dual face with a coefficient $(-1)^m$ (independently of the insertion of the 3d angle observable, see expression \eqref{int area}).
The result of \eqref{conj class} is now modified and the relevant quantity becomes, for $f_1$:
\beq \label{int n 3d angle}
\int_{\SU(2)}dn\ D^{(j)}_{ll}\bigl(n\mone g n\bigr)\ D^{(1)}_{0A}(n\mone) = -(-1)^{j-l}\begin{pmatrix} j & j & 1 \\ -l & l & 0\end{pmatrix} \sum_{a,b} (-1)^{j-a}(-1)^{1-A} \begin{pmatrix} j & j & 1 \\ -a & b & -A \end{pmatrix} D^{(j)}_{ab}(g)
\ee
The quantity for $f_2$ can be obtained using $D_{A0}^{(1)}(n)=(-1)^A D^{(1)}_{0-A}(n\mone)$. Because of the sum over the orientation in \eqref{insertion 3d angle}, we have to evaluate the properties of \eqref{int n 3d angle} under the change $l\rightarrow -l$. Given the properties of the 3jm-symbols, one sees that only the sign is changed. In contrast with $Z_{\mathrm{BF}}$, it implies that the variable $m_f$, dual to the deficit angle $\theta_f$, has to be odd for $f_1$ and $f_2$, and then that the representations $j_1$ and $j_2$ are in $\N+\f{1}{2}$. Notice that the first term in \eqref{insertion 3d angle}, corresponding to $m=0$, does not contribute since the 3jm-symbols $\begin{pmatrix} j & j & 1 \\ 0 & 0 & 0\end{pmatrix}$ are zero for $j\in\N$. Like in pure BF theory, the sum over the angular momenta $m_1$ and $m_2$ can be explicitly performed thanks to the special form of the 3jm-symbols. Indeed, $\sum_{k=1/2}^j (-1)^{j-k} \begin{pmatrix} j & j & 1 \\ -k & k & 0\end{pmatrix} = \f{d_j}{8}\sqrt{\f{d_j}{j(j+1)}}$.

\begin{figure} \begin{center}
\includegraphics[width=10cm]{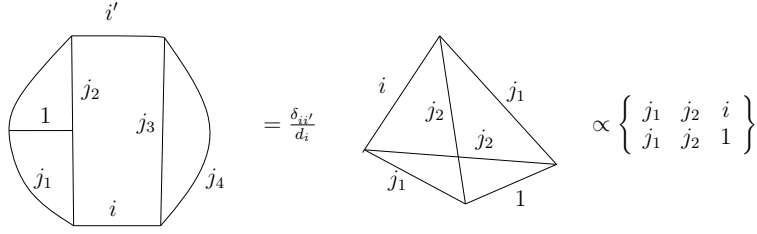}
\caption{ \label{3dinsertion} The left figure shows the graph weighting the tetrahedron $t$  and resulting from the insertion of the 3d angle between $f_1$ and $f_2$ in $t$. In contrast with pure BF theory, it does not simply consist of the contraction of the intertwiners coming from the integrations of $g_1$ and $g_2$ because of the insertion of a link in the spin 1 representation between the links of $f_1$ and $f_2$. This graph reduces to the evaluation of a 6j-symbol if the pairing is conveniently chosen, displaying the role of the intertwiners in the quantization of the 3d angle. }
\end{center}
\end{figure}

The indices $a_1,a_2$ and $b_1,b_2$ of the matrix elements $D^{(j_1)}_{a_1b_1}(g_1)$ and $D^{(j_2)}_{a_2b_2}(g_2)$ are, after integration over the holonomy degrees of freedom, carried by two intertwiners, with different internal representations $i$ and $i'$ taking place at the tetrahedron $t$. In the absence of the observable insertion, one has instead of the 3jm-symbol of \eqref{int n 3d angle} a Kronecker delta, $\delta_{a b}$. Now, the contraction of these two intertwiners is not made according to the orthogonality relation. A link in the spin 1 representation is inserted between the links $j_1$ and $j_2$ of the two faces, as shown in figure \ref{3dinsertion}. However, if the pairing is well chosen, the equality $i=i'$ still holds, and the graph reduces to that of a 6j-symbol. Assuming without loss of generality that $f_1$ and $f_2$ are similarly oriented and gathering the numerical coefficients, the insertion of $\cos\phi^t_{ff'}$ translates into the insertion into $Z_{BF}$ written as a sum over spin foams of the following quantity:
\beq
\f{(-1)^{j_1+j_2+i_t}}{16}\,\sqrt{\f{d_{j_1}^3\,d_{j_2}^3}{j_1(j_1+1)j_2(j_2+1)}}\ \begin{Bmatrix} j_1 & j_1 & 1 \\ j_2 & j_2 & i_t\end{Bmatrix},\qquad j_1,j_2\,\in \N+\f{1}{2}
\ee
where the quantity into brackets denotes the 6j-symbol represented in figure \ref{3dinsertion} with a tetrahedral graph. The angle between two triangles is thus quantized as a 6j-symbol. The latter not only involves $j_1$ and $j_2$ which contain the areas of $f_1$ and $f_2$, but also the representation $i_t$ of the intertwiner at which the triangles meet. The intertwining data $\{i_t\}$ of $Z_{\mathrm{BF}}$ thus encode the correlations between the directions of triangles. 

\section*{Conclusion}

Precise relations between lattice BF theory, supplemented with the simplicity constraints, and area-angle Regge calculus are given. The expressions of the constraints in area-angle Regge calculus and of the (4d) dihedral angles in terms of the 3d angles take a simple form through a formulation of the gluing relations and cross-simplicity with group elements. The dihedral angles enable to express holonomies in terms of bivectors, and appear as part of a $\u(1)\oplus\u(1)$ discrete connection. We propose two actions (the difference being rather a technical ambiguity) which include the Immirzi parameter. It might be interesting to study some other ambiguities, such as choosing a higher spin representation instead of the fundamental.

Our work leads to a mixed setting, combining variables from both Regge calculus and lattice BF gauge theory, and whose geometric content is crystal-clear. As emphasized in the introduction, we believe that it is important to have a (compactified) Regge-like action for which the introduction of the Immirzi parameter makes sense, in particular while aiming at understanding its nature at the quantum level. The Immirzi dependence here disappears from the classical equations of motion for non-degenerate configurations. The case of degenerate configurations is less clear, but the recent classification results of \cite{barrett fairbairn} are certainly helpful in this regard.

Finally, insertions of areas and 3d angles are converted into spin foam language. As expected, areas are quantized according to the Casimirs of triangle representations. Their quantization in the presence of the simplicity constraints is briefly discussed. The insertion of a 3d angle corresponds to the insertion of a 6j-symbol, whose data are the spins of the two involved triangle representations and the tetrahedral intertwining spin.

Let us briefly argue that the present setting is well suited to study spin foam models from discretized path integrals. We can choose and completely control how cross-simplicity and parallel transport relations are taken into account. Indeed, instead of the strong gluing using delta functions over $\SU(2)$, \eqref{gluing measure spin4}, the parallel transport relations can be inserted with another weight. It should be peaked around the identity, so that the equations of motion give the correct rules, and take the following form:
\beq
\tl{\delta}_\vareps\bigl(n_{ft}\mone\,g_{tt'}\,n_{ft'}\,e^{\f{i}{2}\theta_{fv}\sigma_z}\bigr) = \sum_{j\in\f{\N}{2}} \sum_{m=-j}^j f_{jm}(\vareps)\ \bra j,m\lv \ n_{ft}\mone\,g_{tt'}\,n_{ft'}\,e^{\f{i}{2}\theta_{fv}\sigma_z}\ \rv j,m\ket + \mathrm{c.c.}
\ee
Only diagonal elements of the representation matrices are considered, the integrals over the angles $\theta_{fv}^\pm$ and $\psi_{ft}$ leading anyway to this situation \cite{lagrangian SF}. The strong gluing corresponds to the choice $f_{jm}=2j+1$. $\vareps$ is a generic variable which suggests to consider families of models. For instance, we may take $\tl{\delta}_\vareps$ to be a one-parameter family of Gaussians of width $\vareps$:
\beq
\tl{\delta}_\vareps(k)=\f{1}{N}e^{\f{\vareps}{2}\tr^2(k\,\vec{\sigma})},\ \text{with}\qquad f_{jm}(\vareps) = \f{1}{N'}\Bigl(I_j(\vareps) - I_{j+1}(\vareps)\Bigr)
\ee
where $N$ and $N'$ are normalisations and $I_j$ is a modified Bessel function, defined by $I_j(z) = i^jJ_j(-iz)$. Notice that this is quite similar to the setting used in semi-classical computations of correlation functions from spin foam models \cite{graviton}: one inserts a boundary state which assigns a mean geometry to the boundary of the spacetime region under consideration. The above functions precisely assign, as we have shown, mean dihedral angles between the boundary tetrahedra for each 4-simplex. The spin foam models resulting from different choices of $\tl{\delta}$ will all exhibit similar structures, since spin foams are based on the integrations of the variables $n_{ft}$ and $g_{vt}$. From this perspective, the coefficients $f_{jm}$ may be seen as a measure ambiguity, exactly like that affecting the definition of the EPRL model. While the differences between two spin foam models differing by their sets of coefficients $f_{jm}$ may be rather arcane, computing the corresponding functions $\tl{\delta}$ would teach us precisely about these differences. A crucial issue is then obviously the ability to resum these coefficients after the integrations. However, some interesting choices of functions can prevent us from such a challenge, such as coefficients of the form: $f_{jm}\propto \delta_{j,\lv m\rv}$, as will be shown in \cite{lagrangian SF}. In any case, the possibility of easily tuning the fluctuations around the gluing relations is a promising tool towards the understanding of the content of spin foam models. The same approach can be applied to cross-simplicity, as written in the third equation of \eqref{gluing and simplicity}. The resulting spin foam models will be presented elsewhere \cite{lagrangian SF} and in particular the way the new spin foam models for quantum gravity arises.

\section*{Acknowledgements}

The author thanks Etera R. Livine for his availibility to discuss spin foams. He also thanks Simone Speziale and Matteo Smerlak for an instructive discussion about this specific work.


\end{document}